# Climate Modelling of Mass–Extinction Events: A Review

**Georg Feulner**

Potsdam-Institut für Klimafolgenforschung (PIK), P.O. Box 60 12 03, D–14412 Potsdam, Germany
E-mail: feulner@pik-potsdam.de

**Abstract:** Despite tremendous interest in the topic and decades of research, the origins of the major losses of biodiversity in the history of life on Earth remain elusive. A variety of possible causes for these mass-extinction events have been investigated, including impacts of asteroids or comets, large-scale volcanic eruptions, effects from changes in the distribution of continents caused by plate tectonics, and biological factors, to name but a few. Many of these suggested drivers involve or indeed require changes of the Earth's climate, which then affect the biosphere of our planet causing a global reduction in the diversity of biological species. It can be argued, therefore, that a detailed understanding of these climatic variations and their effects on ecosystems are prerequisites for a solution to the enigma of biological extinctions. Apart from investigations of paleoclimate data of the time periods of mass extinctions, climate-modelling experiments should be able to shed some light on these dramatic events. Somewhat surprisingly, however, only few comprehensive modelling studies of the climate changes associated with extinction events have been undertaken. These studies will be reviewed in this paper. Furthermore, the role of modelling in extinction research in general and suggestions for future research are discussed.

## 1. A Brief Introduction to Mass Extinctions in Earth's History

Mass extinction events are certainly among the most dramatic incidents in the history of life on Earth. They are defined as comparatively short intervals of geological time characterised by the disappearance of more than one geographically widespread higher taxon[1] (Bambach, 2006). Almost 20 such events have been identified during the Phanerozoic (i.e. the last 542 million years of Earth's history), with five particularly dramatic events[2]. These "big five" extinction events are the End Ordovician extinction about 445 million years ago, the Late Devonian about 375 million years ago, the famous End Permian (Permian-Triassic or P/T) event 251 million years ago, the End Triassic roughly 200 million years ago, and the End Cretaceous (Cretaceous-Tertiary or K/T) event 65 million years ago, which marks the end of the dinosaurs (Raup & Sepkoski, 1982). Note, however, that not all extinction events are necessarily characterised by elevated extinction intensities: low species origination levels have contributed to some of these turnovers in the biological record (Bambach et al., 2004).

---

[1] A taxon represents one particular rank in the hierarchical system of biological taxonomy; in ascending order, these ranks are species, genus, family, order, class, phylum (in plants: division), and kingdom. 'Higher taxa' are all ranks above the species level.

Understanding the reasons for species mass extinctions is not only of academic interest, but particularly important for two reasons. First, most of the species that have ever lived on this planet have disappeared. Extinction is therefore a key driver in the evolution of life on Earth, especially when biological taxa dominating parts of the biosphere are replaced by the survivor from other taxa, as beautifully illustrated by the rise of the mammals after the disappearance of the dinosaurs. Secondly, the knowledge of the mechanisms giving rise to extinction events in the Earth's past may help mankind in managing the present-day loss of biological diversity.

A variety of different causes have been suggested for the extinction events documented in the fossil record (Ward, 2007). These will be briefly summarised below, starting with extraterrestrial mechanisms for changes in the biological diversity on Earth.

Impacts of small bodies from the solar system (asteroids or comets) might affect the local biosphere directly by causing blast damage, earthquakes, fires or tsunamis, and the whole globe by eject pulverised rock, sulphate aerosols and soot into the stratosphere, which might block sunlight for several years (Toon *et al.*, 1997). This hypothesis has, of course, been proposed in particular for the K/T event (Alvarez *et al.*, 1980), and this, in fact, is the only extinction event where such a connection could be convincingly established, although not consensus has been reached so far.

Other astronomical events might influence the Earth's biosphere as well. High-energy radiation from violent explosions of astronomical objects ranges among the more exotic hypotheses for the causes of mass extinctions. A nearby supernova explosion could have triggered a destruction of the Earth's ozone layer, resulting in harmful ultraviolet radiation reaching the surface (Ellis & Schramm, 1995). A Gamma Ray Burst within the Milky Way galaxy could cause similar damage (Melott *et al.*, 2004), but for both drivers the expected frequency of their occurrence and the effects on the atmosphere and the biosphere are still uncertain.

The causes for mass extinctions, however, could also be found within the Earth system itself. One particular prominent terrestrial mechanism for mass extinctions are large-scale volcanic eruptions (Wignall, 2005). Large eruptions of low-viscosity basaltic magma can cover large areas of land and lead to short-term atmospheric cooling through the production of aerosols and long-term climatic warming through the emission of greenhouse gases. Prominent examples for flood-basalt provinces associated with these eruptions are the Siberian Traps, the Central Atlantic Magmatic Province, and the Deccan Traps in India. Indeed, there appears to be an intriguing correlation between many flood-basalt provinces and extinction events, in particular for the three most severe extinctions, the Permian-Triassic, the Triassic-Jurassic, and the Cretaceous-Tertiary events, which coincide with the three large basaltic provinces mentioned above. There are, however, also examples for eruptions without extinctions, and extinctions without eruptions, so the role of these large-scale volcanic eruptions remains unclear.

---

[2] Note that some authors reserve the term 'mass extinction' for these five events while referring to the other events as 'extinctions'.

Changes of the Earth's climate, most notably rapid global cooling or warming, certainly have the power to affect the biosphere. Indeed, many periods of biodiversity loss in the history of life are correlated with changes in the climate (Twitchett, 2006). Climatic changes could be caused by both by most of the drivers already discussed, but also by changes in the configuration of continents due to the motion of tectonic plates, by variations in the solar radiation received by Earth or by changes of the chemical constituents of the atmosphere (in particular green-house gases). These changes in forcing are then amplified by the positive feedbacks in the climate system (Saltzmann, 2002).

Furthermore, the complex nature of the climate system allows for the possibility of abrupt transitions between different climate states, often requiring only a small change in forcing. Such abrupt climate shifts could lead to impacts on the biosphere. Indeed, some correlation between rapid climatic transitions in Earth's history and the paleontological record seems to exist (Crowley & North, 1988). These shifts in climate could be brought about by any change in forcing, including impacts, volcanoes, plate tectonics, or insolation, and could be an integral element of the connection between the drivers and the biosphere impacts. This further highlights the importance of understanding and modelling the climatic changes associated with the periods of mass extinctions in Earth's history. The past and future of climate-modelling studies of periods of mass extinction are the focus of this review paper.

This paper is organised as follows. Section 2 gives a very short introduction to the climate system and describes the different types of numerical models used in simulating Earth's climate. The climatic effects involved with various origins of mass extinction events are summarised in Section 3, before climate-modelling studies of mass-extinction events are reviewed in Section 4. Section 5, finally, discusses the fundamental requirements of climate models used in extinction research and makes suggestions for future research on extinction events in the history of life on Earth.

## 2. The Climate System and the Hierarchy of Climate Models

Before the climatic effects of different causes of extinctions and previous climate-modelling studies of extinction events are reviewed, we give a brief overview over the various types of climate models. The Earth's climate is primarily determined by the boundary conditions imposed on the Earth system. Among these, the flux of solar radiation received by our planet is most prominent, but other factors like the distribution of continents, the topography of the Earth, the bathymetry of the ocean basins, the availability of chemical elements and many others play an important role as well. Changes of the boundary conditions, which are usually referred to as climate forcings, lead to changes in the Earth's climate.

The climate system is composed of a number of different components: The atmosphere, the hydrosphere (comprising oceans, rivers, lakes), the biosphere (living beings on Earth), the cryosphere (snow, sea ice, glaciers, ice sheets and shelves), the pedosphere (Earth's continental surface) and the lithosphere (Earth's crust and upper mantle). All these components typically are non-linear systems with characteristic time-scales, and all components interact with each other in a non-linear manner (Saltzmann, 2002). Climate models describe these interactions in mathematical terms, which try to reproduce the systems behaviour. Owing to the complexity of the system, these coupled equations have to be solved numerically.

An assessment of existing climate-modelling studies of extinction events requires a basic understanding of the various types of climate models. There are a variety of different climate models with a large range of complexities, which can largely be grouped into a hierarchy of models. The simplest models are energy-balance models (or EBMs for short) which solve the energy balance equation for the whole atmosphere (North & Stevens, 2006). Spatial variations within the atmosphere can be approximated by simplified equations for the vertical and/or horizontal energy transport.

The other end of the hierarchy is formed by models, which try to solve the dynamical equations for the three-dimensional atmosphere on short timescales, usually coupled to a fully three-dimensional model of the oceans. These are the so called general circulation models (often referred to as GCMs), which are, for example, the class of models mostly used in predicting the future climate under the influence of anthropogenic greenhouse-gas emissions (Randall et al., 2007). The major disadvantage of these fully coupled climate models running at high spatial and temporal resolutions is that they are computationally rather expensive and are thus usually used for simulating the climate over time intervals of a few hundred years at most.

In between these extremes are Earth-system models of intermediate complexity (or EMICs), which are much faster than traditional general circulation models due to non-three-dimensional treatment of either the ocean or the atmosphere, lower spatial and temporal resolution and simplified governing equations (Claussen et al., 2002). A special class of EMICs reduces complexity by describing the dynamical behaviour of the atmosphere on large scales in a statistical manner. Intermediate-complexity models are able to simulate the climate over extended periods of time or run many simulations of a shorter time span, so they are ideal tools for studies of paleoclimate, long-term effects of global warming, or sensitivity studies involving ensemble simulations.

Note that the more complex general circulation models are not necessarily superior to intermediate complexity models; often intermediate-complexity models emulate more components of the climate system and their fundamental interactions than general circulation models.

## 3. The Role of Climate Change in Mass-Extinction Events

It is obvious from the various mechanisms proposed for the reduction of biodiversity during periods of mass extinction that many hypotheses involve changes to the Earth's climate. Indeed, some of the primary causes like impacts from space or large-scale volcanic eruptions are primarily local events, which rely on the climate system to make them truly global changes being able to affect the biosphere of large parts of the Earth's surface. Table 1 summarises the causes of extinctions most widely discussed (and briefly described in Section 1) as well as their climatic impacts and the typical timescale on which they affect the climate. Note that beyond their climatic impacts most of these drivers do have more direct impacts on the local biosphere as well, which are beyond the scope of this paper.

| Cause of extinction | Climate impact | Timescale | References |
|---|---|---|---|
| Comet or asteroid impact | Sulphate aerosols, pulverised rock and soot in the stratosphere block incoming sunlight | Years | (Alvarez et al., 1980; Covey et al., 1994; Pope et al., 1994) |
| Large-scale volcanism | Short-term: Stratospheric aerosols cause cooling | Years | (Self, 2005; Wignall, 2005) |
| | Long-term: Carbon dioxide emissions cause global warming | $10^5$ years | |
| Climate forcings in general (plate tectonics, solar radiation, chemistry) | Gradual or abrupt climate change (temperature, precipitation patterns, glaciations, ocean currents, oxygen content of oceans, methane release) | $10^2$-$10^5$ years | (Crowley & North, 1988) |
| Supernova, Gamma Ray Burst | Changes in atmospheric chemistry, reduction/destruction of ozone layer | Years? | (Ellis & Schramm, 1995; Melott et al., 2004) |

Table 1: Summary of climatic effects of various extinction drivers.

Hence paleoclimate data and climate modelling are essential for understanding the changes within the Earth system giving rise to severe species extinctions. Furthermore, they can provide information about the geographic, temporal and seasonal distribution of these changes, which can then be compared to paleontological data on the distribution of species loss in various parts of the Earth, as a function of time, and within different branches of the tree of life. These could serve as 'fingerprints' of the various extinction drivers and thus make it possible to distinguish between different theories on the origins of one particular mass extinction event[3]. It will become clear from the following review of modelling studies that much remains to be done, both in terms of better paleodata and more comprehensive climate models applied to the problem of mass extinctions.

## 4. Climate Modelling of Species Extinction Events – a Review

In this Section, existing studies of mass-extinction events using climate models will be reviewed. Naturally, most investigations focus on the two most prominent periods of biodiversity loss, the P/T and K/T events 251 and 65 million years ago, respectively. We concentrate on the limited number of studies, which focus on the extinction event itself; we note, however, that many more climate-modelling studies of the climate during the geological periods before or after extinction events can be found in the scientific literature.

*The Permian-Triassic Extinction*

The extinction at the Permian-Triassic boundary about 251 million years ago was the most severe extinction event within the last half billion years: More than 90% of all marine species and about two thirds of all land-dwelling species became extinct. Paleodata show that the event was associated with a warm climate and anoxic oceans (White, 2002). Thus many modelling experiments focused on the late-Permian oceans (see Winguth & Maier-Reimer, 2005, for a review). Hotinski *et al.* (2001) and Zhang *et al.* (2001) used uncoupled three-dimensional ocean models with simplified biogeochemistry and prescribed boundary conditions to test this hypothesis, yielding somewhat conflicting results (see also Hotinski *et al.*, 2002; Zhang *et al.*, 2003). Winguth *et al.* (2002) and Winguth & Maier-Reimer (2005) studied the sensitivity of ocean circulations to changes in greenhouse-gas concentrations, strong freshwater perturbation and to massive methane release using an energy-balance model of the atmosphere coupled to an ocean general circulation model. They find that most model simulations result in strong deep-sea circulation patters with relatively high oxygen content in the oceans, thus not supporting the hypothesis of ocean anoxia causing the End-Permian extinction event. However, all these studies suffer from incomplete knowledge of the bathymetry of the Permian oceans and lack a realistic representation of the dynamical behaviour of the atmosphere.

---

[3] Note that this is particularly important for testing the hypothesis of periodic extinction events (Bailer-Jones, 2009).

Most recently, Kiehl & Shields (2005) were able to show that a fully coupled ocean-atmosphere general circulation model for the late-Permian geography and with high levels of carbon dioxide is able to reproduce paleodata indicating high temperatures over the Pangaea supercontinent and a weak overturning circulation yielding anoxic oceans. Excessively high land temperatures are hypothesised to be the reason for the loss of species on the continent. No sensitivity studies, however, were possible with their computationally rather expensive model, and – apart from the high concentrations of carbon dioxide speculated to originate from the Siberian Traps – no direct connection to the ultimate cause of the P/T event is accomplished.

*The Triassic-Jurassic Extinction*

Although not quite as dramatic as the end-Permian extinction, the biodiversity loss at the Triassic-Jurassic boundary about 200 million years ago ranges among the 'big five' extinctions in the history of life on Earth. It has been suggested that the eruption of the Central Atlantic Magmatic Province (CAMP) and thus enhanced carbon dioxide concentration in the atmosphere could be responsible for the extinction event at the end of the Triassic (Wignall, 2005).

Huynh & Poulsen (2005) used a low-resolution atmosphere and ocean general circulation model with somewhat simplified vegetation patterns to simulate the climate at the time of the extinction event. In a series of sensitivity experiments with carbon dioxide levels between twice and eight times the pre-industrial concentrations they demonstrate the existence of extreme conditions. On land, the climate is characterised by an increase of hot and dry days and enhanced seasonality, while the oceans suffer from decreasing overturning and low oxygen concentrations, leading to stress to both continental and marine environments.

*The Cretaceous-Tertiary Extinction*

The famous mass extinction at the Cretaceous-Tertiary boundary approximately 65 million years before present marks the end of the dinosaurs. Two competing scenarios have been invoked to explain this dramatic event: an impact from space or the giant volcanic eruptions giving rise to the Deccan Traps in India. Simulations of the climate changes related to the impact is complicated by the fact that the respective opacities and atmospheric lifetimes of dust, sulphate aerosols and soot are still uncertain (Kring, 2007). Many modelling studies are based on radiative transfer and atmospheric chemistry models (e.g., Pope *et al.*, 1997), with only very few more sophisticated climate simulations to be found in the literature.

Covey *et al.* (1994) used a low-resolution atmospheric general circulation model in combination with a rather simple ocean model – unfortunately run on present-day geography – to simulate the climatic effects of a large dust load produced by the impact of a comet or small asteroid. In their simulations, they find a strong continental cooling of about 13 degrees centigrade for a few days, after which the temperatures recover to 6 degrees below normal by the end of the first year after the impact. This cooling is accompanied by a breakdown of the hydrological cycle, with a globally averaged decrease in precipitation of more than 90% for several months and about 50% after one year.

## 5. Discussion, Conclusions and Outlook

From the few climate-modelling studies of extinction events found in the scientific literature, it is obvious that many details of the chain linking proposed extinction causes to the biodiversity loss merit further investigation. Climate models clearly play an integral part in this, but should meet a number of conditions:

- To provide a reasonable representation of the geographic distribution of climate impacts, general circulation models or models of intermediate complexity should be used (rather than the overly simplistic energy-balance models). For intermediate-complexity models in particular, a three-dimensional ocean component is clearly important, and the atmospheric component coupled to the ocean model should provide both sufficient resolution and coverage of the basic dynamic patterns. Indeed, given the uncertainties in the paleodata and the need for sensitivity experiments, intermediate-complexity Earth-system models meeting these requirements may be the ideal tools for studies of climatic transitions associated with extinction events.
- Because of the importance of the distribution of continents and of ocean currents for the Earth's climate, the models should be run on a realistic geography and topography for the time of the extinction event.
- Feedbacks between climate and the biosphere are an important factor in the climate system, so approximations of the continental vegetation and the marine carbon cycle should be included in the modelling experiments, provided that sufficient paleontological data are available. Other crucial components of the climate system (e.g. sea ice and continental ice sheets) should be included as well.
- For particular extinction causes under investigation (e.g. astronomical impacts or volcanic eruptions), the effects of these drivers on the atmosphere should be realistically included into the climate models. For extinction events with competing mechanisms suggested as their cause, climate-modelling experiments should focus on possible differences between these scenarios, which might allow an assessment of the respective contribution of the different drivers. Ideally, effects on the atmospheric greenhouse-gas or aerosol budget should be linked to the driver, although sensitivity studies using a range of greenhouse-gas concentrations derived from paleoclimate data may have to serve as an intermediate step.
- Some suggested extinction drivers involve changes in the stratospheric chemistry, requiring a reasonable approximation of these processes linked to or included into the models.

Considering these basic requirements, it becomes obvious that only very few of the available modelling studies of the climatic changes associated with extinction events in the Earth's history come close to meeting most of these conditions. Fulfilling these requirements should be the guideline for future work on the climatic changes associated with species extinctions.

Furthermore, empirical data are unavailable for some important variables needed for the models, and improved paleodata might be necessary to ensure a realistic representation of the climate. Of course, some uncertainties about these and other variables in the Earth's past are likely to remain, which will negatively affect the accuracy of the modelling studies, especially for the more ancient extinction events. Nevertheless, improved climate modelling experiments of extinction events are urgently needed to assess the impact of various suggested causes on different parts of the Earth's system, hopefully providing an answer to the intriguing puzzle of the major periods of biodiversity loss in the history of our planet.

**Acknowledgements**

This work has benefited greatly from discussions at the ESLAB 2008 conference on *Cosmic Cataclysms and Life*, but also at the Astrobiology Science Conference (AbSciCon) 2008 and the 8$^{th}$ European Workshop on Astrobiology (EANA'08). Travel funding for these meetings was partly provided by the European Space Agency (ESA), the Deutsche Forschungsgemeinschaft (DFG, grant FE 1060/1–1), and the European Astrobiology Network Association (EANA), respectively. This financial support is gratefully acknowledged.

**References:**

Alvarez, L. W., W. Alvarez, F. Asaro & H. V. Michel (1980). "Extraterrestrial Cause for the Cretaceous-Tertiary Extinction." Science 208(4448): 1095-1108.

Bailer-Jones, C. (2009). "The evidence for and against astronomical impacts on climate change and mass extinctions: A review." International Journal of Astrobiology in press.

Bambach, R. K. (2006). "Phanerozoic biodiversity mass extinctions." Annual Review of Earth and Planetary Sciences 34: 127-155.

Bambach, R. K., A. H. Knoll & S. C. Wang (2004). "Origination, extinction, and mass depletions of marine diversity." Paleobiology 30(4): 522-542.

Claussen, M., L. A. Mysak, A. J. Weaver, M. Crucifix, T. Fichefet, M. F. Loutre, S. L. Weber, J. Alcamo, V. A. Alexeev, A. Berger, R. Calov, A. Ganopolski, H. Goosse, G. Lohmann, F. Lunkeit, Mokhov, II, V. Petoukhov, P. Stone & Z. Wang (2002). "Earth system models of intermediate complexity: closing the gap in the spectrum of climate system models." Climate Dynamics 18(7): 579-586.

Covey, C., S. L. Thompson, P. R. Weissman & M. C. MacCracken (1994). "Global climatic effects of atmospheric dust from an asteroid or comet impact on Earth." Global and Planetary Change 9(3-4): 263-273.

Crowley, T. J. & G. R. North (1988). "Abrupt Climate Change and Extinction Events in Earth History." Science 240(4855): 996-1002.

Ellis, J. & D. N. Schramm (1995). "Could a nearby supernova explosion have caused a mass extinction?" Proceedings of the National Academy of Sciences of the United States of America 92(1): 235-238.

Hotinski, R. M., K. L. Bice, L. R. Kump, R. G. Najjar & M. A. Arthur (2001). "Ocean stagnation and end-Permian anoxia." Geology 29(1): 7-10.

Hotinski, R. M., L. R. Kump & K. L. Bice (2002). "Comment on "Could the Late Permian deep ocean have been anoxic?" by R. Zhang et al." Paleoceanography 17(4): 2.

Huynh, T. T. & C. J. Poulsen (2005). "Rising atmospheric CO2 as a possible trigger for the end-Triassic mass extinction." Palaeogeography Palaeoclimatology Palaeoecology 217(3-4): 223-242.

Kiehl, J. T. & C. A. Shields (2005). "Climate simulation of the latest Permian: Implications for mass extinction." Geology 33(9): 757-760.

Kring, D. A. (2007). "The Chicxulub impact event and its environmental consequences at the Cretaceous-Tertiary boundary." Palaeogeography, Palaeoclimatology, Palaeoecology 255(1-2): 4-21.

Melott, A. L., B. S. Lieberman, C. M. Laird, L. D. Martin, M. V. Medvedev, B. C. Thomas, J. K. Cannizzo, N. Gehrels & C. H. Jackman (2004). "Did a gamma-ray burst initiate the late Ordovician mass extinction?" International Journal of Astrobiology 3(01): 55-61.

North, G. R. & M. J. Stevens (2006). Energy-balance climate models. In: Frontiers of Climate Modeling. J. T. Kiehl and V. Ramanathan (Eds.). Cambridge, Cambridge University Press: 52-72.

Pope, K. O., K. H. Baines, A. C. Ocampo & B. A. Ivanov (1994). "Impact winter and the Cretaceous/Tertiary extinctions: Results of a Chicxulub asteroid impact model." Earth and Planetary Science Letters 128(3-4): 719-725.

Pope, K. O., K. H. Baines, A. C. Ocampo & B. A. Ivanov (1997). "Energy, volatile production, and climatic effects of the Chicxulub Cretaceous/Tertiary impact." J. Geophys. Res. 102(E9): 21645-21664.

Randall, D. A., R. A. Wood, S. Bony, R. Colman, T. Fichefet, J. Fyfe, V. Kattsov, A. Pitman, J. Shukla, J. Srinivasan, R. J. Stouffer, A. Sumi & K. E. Taylor (2007). Climate Models and Their Evaluation. In: Climate Change 2007: The Physical Science Basis. Contribution of Working Group I to the Fourth Assessment Report of the Intergovernmental Panel on Climate Change. S. Solomon, D. Qin, M. Manning *et al* (Eds.). Cambridge, Cambridge University Press.

Raup, D. M. & J. J. J. Sepkoski (1982). "Mass Extinctions in the Marine Fossil Record." Science 215(4539): 1501-1503.

Saltzmann, B. (2002). Dynamical Paleoclimatology - Generalized Theory of Global Climate Change. San Diego, Academic Press.

Self, S. (2005). Effects of volcanic eruptions on the atmosphere and climate. In: Volcanoes and the Environment. J. Martí and G. G. J. Ernst (Eds.). Cambridge, Cambridge University Press: 152-174.

Toon, O. B., K. Zahnle, D. Morrison, R. P. Turco & C. Covey (1997). "Environmental Perturbations Caused by the Impacts of Asteroids and Comets." Rev. Geophys. 35(1): 41-78.

Twitchett, R. J. (2006). "The palaeoclimatology, palaeoecology and palaeo environmental analysis of mass extinction events." Palaeogeography Palaeoclimatology Palaeoecology 232(2-4): 190-213.

Ward, P. D. (2007). Mass extinctions. In: Planets and Life. W. T. Sullivan III and J. A. Baross (Eds.). Cambridge, Cambridge University Press: 335-354.

White, R. V. (2002). "Earth's biggest 'whodunnit': unravelling the clues in the case of the end-Permian mass extinction." Philosophical Transactions of the Royal Society of London. Series A: Mathematical, Physical and Engineering Sciences 360(1801): 2963-2985.

Wignall, P. B. (2005). Volcanism and mass extinctions. In: Volcanoes and the Environment. J. Martí and G. G. J. Ernst (Eds.). Cambridge, Cambridge University Press: 207-226.

Winguth, A. M. E., C. Heinze, J. E. Kutzbach, E. Maier-Reimer, U. Mikolajewicz, D. Rowley, A. Rees & A. M. Ziegler (2002). "Simulated warm polar currents during the middle Permian." Paleoceanography 17(4): 1057.

Winguth, A. M. E. & E. Maier-Reimer (2005). "Causes of the marine productivity and oxygen changes associated with the Permian-Triassic boundary: A reevaluation with ocean general circulation models." Marine Geology 217(3-4): 283-304.

Zhang, R., M. J. Follows, J. P. Grotzinger & J. Marshall (2001). "Could the Late Permian deep ocean have been anoxic?" Paleoceanography 16(3): 317-329.

Zhang, R., M. J. Follows & J. Marshall (2003). "Reply to Comment by Roberta M. Hotinski, Lee R. Kump, and Karen L. Bice on "Could the Late Permian deep ocean have been anoxic?"." Paleoceanography 18(4): 1095.